\documentclass[11pt, oneside]{article}   	
\usepackage{geometry}                		
\usepackage{graphicx}
\usepackage{amsmath,amssymb}
\usepackage{natbib}
\setcitestyle{round,authoryear,aysep={},yysep={;}}

\usepackage{sectsty} 
\sectionfont{\sffamily\bfseries\large\underline}
\subsectionfont{\sffamily\bfseries\scshape\normalsize} 
\subsubsectionfont{\rmfamily\bfseries\upshape\normalsize} 

	\usepackage{setspace}
\begin{document}
\title{Risk management of solitary and eusocial reproduction}
\author{Feng Fu\\
{\small Theoretical Biology, Institute of Integrative Biology,} \\
{\small ETH Z\"{u}rich, 8092 Z\"{u}rich, SWITZERLAND} \\
Sarah Kocher\\
{\small Department of Organismic and
Evolutionary Biology, Harvard University, Cambridge, MA 02138, USA} \\
Martin A. Nowak\\
{\small Program for Evolutionary Dynamics, } \\
{\small Department of Organismic and
Evolutionary Biology,} \\
{\small Department of Mathematics, }\\
\small Harvard University, Cambridge, MA 02138, USA}
\date{}					
\maketitle

\abstract{Social insect colonies can be seen as a distinct form of biological organization because they function as superorganisms. Understanding how natural selection acts on the emergence and maintenance of these colonies remains a major question in evolutionary biology and ecology. Here, we explore this by using multi-type branching processes to calculate the basic reproductive ratios and the extinction probabilities for solitary versus eusocial reproductive strategies. In order to derive precise mathematical results, we use a simple haploid, asexual model.  In general, we show that eusocial reproductive strategies are unlikely to materialize unless large fitness advantages are gained by the production of only a few workers. These fitness advantages are maximized through obligate rather than facultative eusocial strategies. Furthermore, we find that solitary reproduction is `unbeatable' as long as the solitary reproductive ratio exceeds a critical value. In these cases, no eusocial parameters exist that would reduce their probability of extinction. Our results help to explain why the number of solitary species exceeds that of eusocial ones: eusociality is a high risk, high reward strategy, while solitary reproduction is better at risk management.}

\clearpage

\newpage

\section{Introduction}

Eusocial behavior occurs when individuals reduce their lifetime reproduction in order to help raise their siblings~\citep{Wilson71, Wilson75}. Eusocial colonies comprise two castes: one or a few reproductive individuals and a (mostly) non-reproductive, worker caste. These societies thus contain a reproductive division of labor where some individuals reproduce and others forage and provision the young~\citep{HB90, EOWilson}. 

Insect species that have reached this remarkable level of social complexity have proven to be vastly successful in nature, and are thought to comprise nearly 50\% of the world's insect biomass~\citep{Wilson71,Wilson90}. Despite the apparent success of this strategy, only 2\% of described insect species are eusocial. Here, we use a mathematical approach to improve our understanding of why such a successful life history strategy so rarely evolves.

Eusociality has originated only 12-15 times within the insects, and perhaps 20-25 times within the entire animal kingdom~\citep{Wilson71, Wilson75}. Most of these evolutionary origins have occurred in the order Hymenoptera (the bees, ants, and wasps) where there has been a single origin in the ants~\citep{Wilson71}, 2-3 origins in the wasps~\citep{Hines_PNAS07}, and 4-5 origins in the bees~\citep{Cardinal_PLoS11,Gibbs_MPE12}. As a result, much focus has been placed on understanding the factors favoring or disfavoring the evolution of eusociality within the social insects~\citep{Hamilton_JTB64,Hamilton_ARES72,Andersson_ARES84,Crespi_Nature92,Crozier_AJE08,Strassmann_Nature11}. 

A large but often overlooked emphasis has been placed on identifying some of the key ecological factors associated with the evolution of eusociality. Several precursors for the origins of eusociality have been proposed based on comparative analyses among social insect species -- primarily the feeding and defense of offspring within a nest~\citep{Andersson_ARES84}. These behaviors are also frequently associated with a subsocial life history, where parents and offspring overlap and occupy the nest simultaneously~\citep{Gadagkar09, Hunt07, Hunt_SBW91, Hunt_JEB12, Gadagkar_JG90, Alok_PNAS12}.   	

In contrast to the relatively few origins of eusociality, losses or reversions back to solitary behavior, which depends on the degree of euscociality, appear to be much more common (a very rough estimate indicates a minimum of 9 losses within the bees -- nearly twice the number of origins; see~\citep{Wcislo_TrEE97,Cardinal_PLoS11}). This suggests that the evolution of eusociality is difficult to both achieve and maintain. Despite this apparent difficulty, some taxa have nonetheless evolved highly complex eusocial behaviors where caste differentiation is determined during development and individuals no longer retain the ability to perform all of the tasks required to reproduce on their own (obligate eusociality)~\citep{Batra_IJE66, Crespi_BE95}. As opposed to facultatively social species where all individuals are capable of reproduction, obligately eusocial species are often described as having crossed a ``point of no return'' where reversion to a solitary life history is no longer possible~\citep{Wilson_PNAS05b}.

Here we calculate the emergence and extinction probabilities of eusocial versus solitary reproductive strategies. We use the haploid version of a simple model of eusociality from \cite{Nowak_Nature10}. In our model, solitary females produce offspring that leave the nest to reproduce individually, while eusocial females produce a mix of offspring that remain in the natal nest and others that leave the nest to establish new, eusocial nests. Those offspring that stay augment the size of the colony. We assume that the key reproductive parameters (oviposition rate, successful raising of the young, and expected lifespan of reproductive females) increase with colony size.

Our focus is not on explaining the origins of eusociality, which has already been modeled in~\cite{Nowak_Nature10}. Instead, we attempt to explain the rarity of this trait by calculating the extinction probabilities of solitary versus eusocial reproduction in a stochastic model based on the theory of branching processes~\citep{Harris02, Antia_Nature03}. We calculate the probability that a lineage starting from a single, solitary individual faces extinction, and we compare this with the extinction probability of a lineage that starts with a single, eusocial individual. 

We find that eusociality is typically a riskier strategy than solitary reproduction unless eusocial colonies can benefit from a large reproductive advantage with very small colony sizes. We also find that, when it is available, obligate eusociality is a more successful strategy than facultative eusociality. These results suggest that the evolution of eusociality is difficult because it requires a high reproductive payoff generated by the immediate production of a small number of workers. These findings can explain both why origins of eusociality are rare but hugely successful and why reversions to solitary behavior also occur: compared to solitary reproduction, eusociality is a high risk and high reward strategy.

\section{Material and methods}

Following the deterministic model of eusociality studied in \cite{Nowak_Nature10}, here we consider a corresponding stochastic version for the case of asexual reproduction. This model-based approach can be used to evaluate the probability that a single, eusocial foundress in a population of solitary females  (\emph{e.g.}, a eusocial mutant) will successfully give rise to a eusocial lineage. We explore this by varying three main parameters: birth rate, $b$, death rate, $d$, and offspring dispersal probability, $q$. We relate these variables in two major ways: through the basic reproductive ratio, $R_0$, and the extinction probability, $p$, of a given lineage. These parameters are described in detail below.

A solitary individual reproduces at rate $b_0$, and dies at rate $d_0$. Each offspring of a eusocial queen stays with their mother with probability $q$; otherwise she leaves to establish another new colony with probability $1 - q$. A eusocial queen with colony size $i$ reproduces at rate $b_i$ and dies at rate $d_i$. $b_i$ and $d_i$  are sigmoid functions of the colony size $i$ with the inflection point being at the eusocial threshold $m$. For simplicity, let us start with assuming $b_i$ and $d_i$ to be step functions of $i$: for $i < m$, $b_i = b_0$ and $d_i = d_0$; for $i \ge m$, $b_i = b$ and $d_i = d$. Thus, eusociality confers an enhanced reproductive rate and a reduced death rate for a queen whose colony size reaches the eusocial threshold ($i\ge m$): $b > b_0$ and $d < d_0$ (henceforth referred to as the ``eusocial advantage''). Later on, we will verify the robustness of our derived results using more general sigmoid functions. 

A single foundress can either die or successfully start a lineage by giving birth to another reproductive individual. Each queen reproduces independently of one another. In light of this, we can use a multi-type branching process to calculate the ultimate \emph{extinction probability}, $p$, of individual founding females. This value represents the chance than a given lineage founded by a single female will go extinct over time $t\to \infty$. The converse of this is the \emph{emergence probability}, $1-p$, or the chance that a lineage founded by a single female will persist over time $t\to \infty$.

\begin{figure}[htbp]
   \centering
   \includegraphics[width=.8\columnwidth]{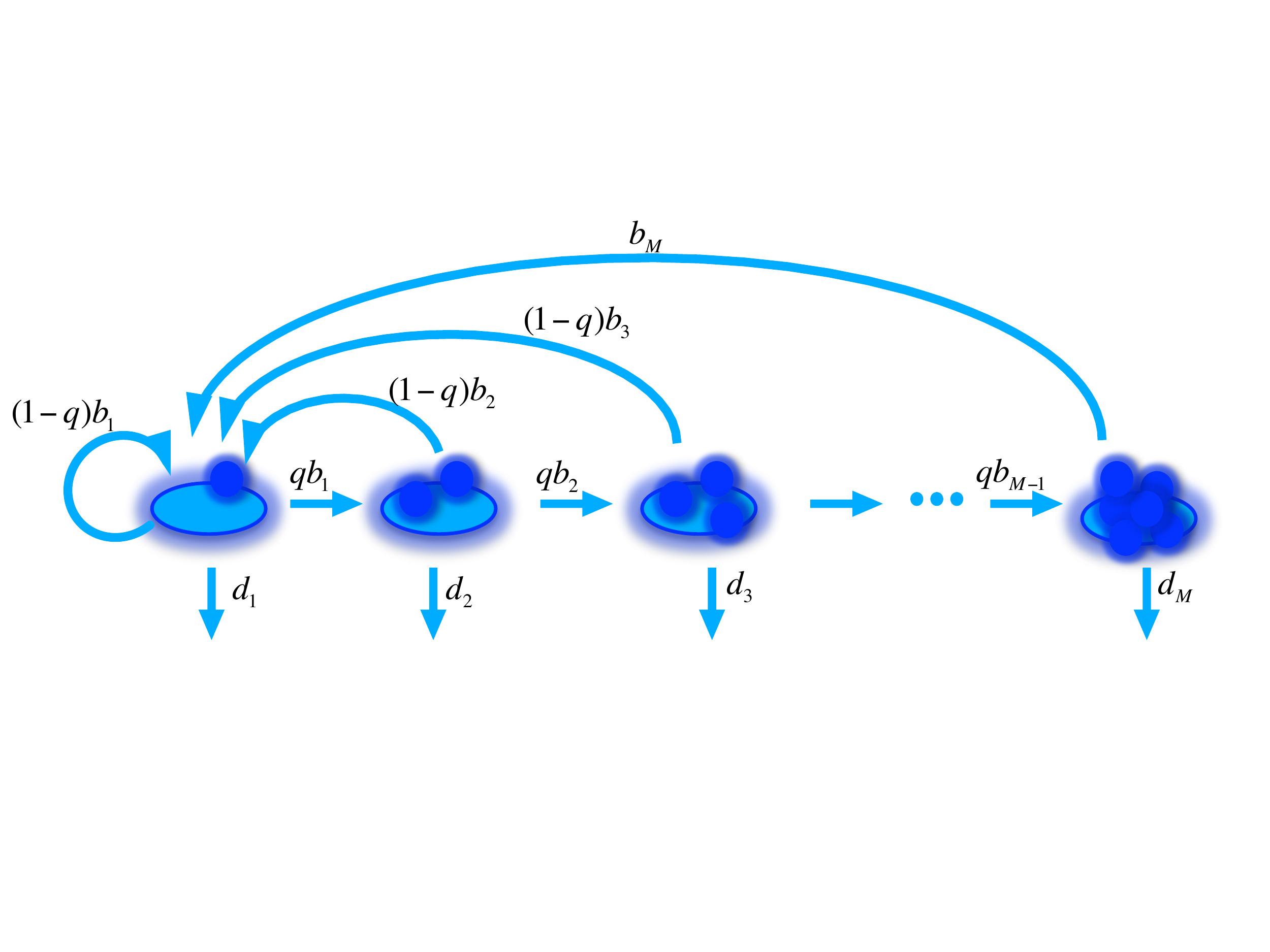} 
      \caption{Schematic illustration of the stochastic branching model. We describe the stochastic emergence of eusociality using a continuous-time multi-type branching process, in which every colony of size $i$ can be regarded as the $i$-th type. We set as $M$ the largest possible size of each eusocial colony, and thus there are total $M$ distinctive types. A colony of size $i$ can transform into the one of size $i+1$ with rate $q b_i$, reproduces a new colony of size $1$ with rate $(1-q) b_i$, or dies with rate $d_i$.}
   \label{fig:example}
\end{figure}

Since the daughters who stay with their mother assist to perform tasks for the colony, the whole colony (consisting of a reproductive queen and non-reproductive workers) can be seen as a sort of superorganism which differs in size and reproductive capacity ($b_i, d_i$). We thus construct the type space using colony size $i$ (Fig. 1). 

For an infinitesimal time interval, $\Delta t$, a colony of size $i$ dies with probability $d_i \Delta t$, increases its size to $i + 1$ with probability $q b_i \Delta t$, and gives rise to a new colony of size $1$ with probability $(1 - q)b_i\Delta t$. When a colony reaches the maximal size $M$, the subsequent offspring are required to leave the colony. Thus the colony of size $M$ gives rise to a new colony of size $1$ with probability $b_M \Delta t$. For simplicity, we do not consider worker mortality. 

Let $f^i(x_1, x_2,\dots, x_M; t)$ denote the joint probability generating function for the number of colonies of size $j = 1, \dots, M$ starting with a single colony of size $i$ at time $t$. Using the backward equation for this continuous-time multiplicative process, we have

\begin{eqnarray}
\frac{d f^i}{dt} & = & d_i + q b_i f^{i+1} + (1 - q) b_i f^1 f^i - (b_i + d_i) f^i, \quad   1\le i < M,
\label{mbp}
\\
\frac{d f^M}{dt} & = & d_M +   b_M f^1 f^M - (b_M + d_M)f^M,   \quad i = M.
\end{eqnarray}

The initial condition is $f^i = x_i$, $i = 1, \dots, M$.

The fixed points of the equations are the ultimate extinction probabilities, $p_i$, starting from a single colony of size $i$. We will give closed form solutions for the extinction probabilities for general cases.

\section{Results}

A population of individuals cannot be sustained without sufficient reproduction~\citep{Holyoak_PLoS14}. This gives rise to the notion of a `\emph{basic reproductive ratio}', $R_0$, or the total expected number of (reproductive) offspring produced by an individual during its lifetime. Simple deterministic models show that supercritical reproduction $R_0 > 1$ is required to sustain a population (\emph{e.g.}, births must exceed deaths). Correspondingly, the condition $R_0 > 1$ also ensures a stochastic supercritical branching process in which the ultimate extinction probability of an individual is not definite but less than one.

\subsection{Solitary reproduction}
Returning to our model, for $q = 0$ (\emph{e.g.}, all offspring disperse) the above Eq.~(1) recovers the case of solitary reproduction. The solitary basic reproductive ratio $R_S$ is

\begin{equation}
R_S = \frac{b_0}{d_0}.
\end{equation}

The extinction probability, $p_0$, of a single solitary individual can be easily found to be
 
\begin{equation}
p_0 = \left\{ \begin{array}{ll}
d_0/b_0, & \quad R_S>1,\\
1, & \quad R_S\le 1.
\end{array}
\right.
\end{equation}

\subsection{Eusocial reproduction with $M = 2$}

Let us consider the simplest possible case of eusociality, with $M = 2$ and $m = 2$. This scenario gives us an intuitive understanding of the results obtained with more general cases. The extinction probabilities satisfy the following equations:

\begin{eqnarray}
d_0 + q b_0 p_2 + (1 - q) b_0 p_1^2 - (b_0 + d_0)p_1 & = & 0,\\
d + b p_1 p_2 - (b + d)p_2 & = & 0.
 \end{eqnarray}

where $p_1$ and $p_2$ denote the extinction probability of a colony of size $1$ and $2$, respectively. Because of the eusocial advantage, we set $b > b_0$ and $d < d_0$. We also assume solitary reproduction is supercritical, that is $b_0 > d_0$. We are primarily interested in the extinction probability of a single eusocial queen, $p_1$, which is a function of $q$:

\begin{equation}
p_ 1 = \frac{b \left(b_0+d_0\right)+(1-q) b_0 d-\sqrt{\left(b \left(b_0+d_0\right)+(1-q) b_0 d\right){}^2-4 (1-q) b_0 b \left(q b_0 d+d_0 \left(b+d\right)\right)}}{2 (1-q) b_0 b}.
\end{equation}

The formula for $p_1$ can be simplified for extremes values of $q$. 

For $q \to 0$, 

\[
p_1 = \frac{d_0}{b_0}.
\]

Note that this is also the ultimate extinction probability of a single ``solitary'' individual. 

For $q \to1$, 

\[
p_1 = \frac{d_0}{b_0+d_0}+\frac{d}{b}.
\]

Furthermore, if the following condition holds,

\begin{equation}
\frac{b}{d} > \frac{b_0}{d_0} (1 + \frac{b_0}{d_0}) 
\label{extp}
\end{equation}

we have $\frac{dp_1(q)}{dq} < 0$ and hence $p_1(q)$ is a decreasing function of $q$, which suggests that a eusocial queen ($0 < q \le 1$) has a greater emergence probability than a solitary one. We also find that the eusocial basic reproductive ratio,

\begin{equation}
R_E = b_0(1-q)/(d_0 + b_0 q) + b b_0 q/[d(d_0 + b_0q)].
\end{equation}

Eusocial $R_E$ is greater than the solitary $R_S$ if

\begin{equation}
\frac{b}{d} > 1 + \frac{b_0}{d_0}.
\label{ro}
\end{equation}

Note that $b_0/d_0 > 1$, the critical $b/d$ above which eusociality has a greater probability of emergence than solitary (Eq.\ref{extp}), is larger than the $b/d$ obtained by simply requiring that eusocial $R_E$ is greater than solitary $R_S$ (Eq.\ref{ro}). This result can help us to understand why a higher eusocial $R_E$ relative to a solitary $R_S$ alone is not sufficient to guarantee that a higher emergence probability for a eusocial queen relative to a solitary female.

\subsection{General case with $M > 2$}

\paragraph{Supercritical condition for eusociality.}
We now turn to more general cases with $M > 2$ and $m \ge 2$. For notational simplicity, let us define $\mu_i = b_i q$ for $1\le i < M$ and $\mu_M = 0$; $\nu_i = b_i (1 - q)$ for $1\le i < M$ and $\nu_M = b_M$. Let us now derive the condition for eusocial reproduction to be supercritical. To do this, let us first obtain the expectation matrix $\mathbf{K}$ for the multi-type branching process as given in Eqs. (1)-(2). The element of $\mathbf{K}$ is given by $K_{ji} = \frac{\partial f_{i}(1,\cdots,1; t)}{\partial x_j}$. Denote by $y_i$ the mean number of colonies of size $i$ and by $\mathbf{y}$ the vector of $[y_1, y_2, \cdots, y_M]$. Then the mean behavior of the branching process is described by 

\begin{equation}
\dot{\mathbf{y}} = \mathbf{K}\mathbf{y},
\end{equation}

which corresponds to the deterministic ordinary differential equations for eusocial reproduction~\citep{Nowak_Nature10}. The supercritical branching process requires $\alpha(\mathbf{K}) > 0$, where $\alpha(\mathbf{K})$ is the spectral abscissa (the largest of the real parts of the eigenvalues) of the matrix $\mathbf{K}$~\citep{Harris02}. 

Before proceeding further, let us define two matrices $\mathbf{F} = \{F_{ij}\}$ and $\mathbf{V} = \{V_{ij}\}$ as follows.

\begin{equation}
F_{ij} = \left\{ \begin{array}{ll}
\nu_i, & \quad i = 1, 1\le j \le M,\\
0, & \quad \mbox{otherwise},
\end{array}
\right.
\end{equation}

\begin{equation}
V_{ij} = \left\{ \begin{array}{ll}
d_i + \mu_i,  & \quad i = j,\\
- \mu_{i-1}, & \quad i = j +1,\\
0, & \quad \mbox{otherwise}.
\end{array}
\right.
\end{equation}

We can see that both $\mathbf{F}$ and $\mathbf{V}$ have clear biological interpretation and are the so-called `next-generation' matrices~\citep{Diekmann_JMB90}.  The entry $F_{ij}$ of matrix $\mathbf{F}$ denotes the rate at which a colony of size $j$ produces new colonies of size $i$. The entry $V_{ij}$ of $\mathbf{V}$ denotes the (influx) rate at which the class of colonies of size $j$ move into the class of colonies of size $i$ ($i \neq j$); the diagonal entry $(i,i)$ of $\mathbf{V}$ denotes the total outflux rate of colonies of size $i$, including the death rate and the rate at which they move into other classes. The inverse of $\mathbf{V}$ is found to be

\begin{equation}
V_{ij}^{-1} = \left\{ \begin{array}{ll}
1/(d_i + \mu_i), & \quad i = j,\\
\prod_{k = j}^{i - 1} \mu_k / \prod_{k = j}^{i } (d_k + \mu_k), & \quad j <i<M, \\
0, & \quad \mbox{otherwise}.
\end{array}
\right.
\end{equation}

It is straightforward to verify the following matrix decomposition:

\[
\mathbf{K} = \mathbf{F} - \mathbf{V} = (\mathbf{F} \mathbf{V}^{-1} - \mathbf{I}) \mathbf{V}.
\]

We obtain the supercritical condition after some algebra~\citep{Diekmann_book}, 

\begin{equation}
\alpha(\mathbf{K}) > 0 \leftrightarrow \rho(\mathbf{F} \mathbf{V}^{-1}) >1,
\end{equation}

where $\rho(\mathbf{A})$ denotes the spectral radius of a matrix $\mathbf{A}$. Therefore, the eusocial basic reproductive ratio, $R_E$, can be defined as $\rho(\mathbf{F} \mathbf{V}^{-1})$, and has a closed-form expression as follows.

\begin{equation}
R_E = \sum_{i = 1}^{M} \frac{\nu_i}{\mu_i}\prod_{j = 1}^i\frac{\mu_j}{d_j + \mu_j}. 
\end{equation}

It is worth noting that the eusocial $R_E$ is exactly the same as the one obtained by applying the next-generation approach to the deterministic Eq.~(4)~\citep{Diekmann_JMB90}. We obtain a supercritical condition for eusociality based on the basic reproductive ratio (`$R_0$'). Eusociality has a nonzero chance of emergence if the branching process is supercritical, i.e., $\alpha(\mathbf{K}) > 0$. This condition is equivalent to $R_E = \rho(\mathbf{F} \mathbf{V}^{-1} ) > 1$. Accordingly, if the eusocial $R_E$ exceeds one, then eusociality can eventually emerge with non-zero probability. 

\paragraph{Emergence probability.}
For general cases with $M > 2$, we can derive a closed form for the emergence probability of a single eusocial queen, $z_1= 1- p_1$, as follows. Let the vector $\mathbf{z} = [z_1, z_2, \cdots, z_M]$, where $z_i$ is the emergence probability of a single colony with size $i$. Let $\boldsymbol{\Lambda} = \text{diag}\{\nu_1,\cdots,\nu_M\}$ and $\mathbf{I} = \text{diag}\{1,\cdots,1\}$ (the $M\times M$ identity matrix). 
Using $z_i = 1 - p_i$, we obtain the following ordinary differential equations for the time evolution of $\mathbf{z}$ after simple algebra:

\begin{equation}
\dot{\mathbf{z}} = \mathbf{K}^T \mathbf{z} - z_1\boldsymbol{\Lambda} \mathbf{z},
\end{equation}

with the initial condition $\mathbf{z}(0) =[1, 1, \cdots, 1]$. The ultimate emergence probability can be given by

\begin{equation}
\mathbf{K}^T \mathbf{z} - z_1\boldsymbol{\Lambda} \mathbf{z} = (\boldsymbol{\Lambda}^{-1}\mathbf{K}^T - z_1 \mathbf{I})\boldsymbol{\Lambda}\mathbf{z} = 0.
\end{equation}

For the supercritical branching process, we have $z_i>0$ for $i = 1, \cdots , M$. This leads to 

\begin{equation}
\det (\boldsymbol{\Lambda}^{-1}\mathbf{K}^T - z_1 \mathbf{I}) = 0.
\end{equation}

It follows immediately that $z_1$ is determined by the spectral abscissa of $\boldsymbol{\Lambda}^{-1}\mathbf{K}^T$. Therefore, we obtain the following closed-form solution for $z_1$,

\begin{equation}
z_1 = \max\{0, \alpha(\boldsymbol{\Lambda}^{-1}\mathbf{K}^T)\}.
\end{equation}

We note that this closed-form $z_1$ agrees perfectly with the results obtained by numerically solving the differential equations (Fig. 2). 

\paragraph{Resilience against extinction: solitary reproduction vs. eusocial reproduction.}
We were interested in whether a eusocial queen can always have a higher emergence probability than a solitary individual given sufficiently large eusocial reproductive advantages. We find that it is always possible to satisfy the condition that the eusocial $R_E$ is greater than the solitary $R_S$. Thus eusocial reproduction is a `high reward' strategy. Counterintuitively, however, it is not always possible for eusociality to have a lower extinction probability than solitary, no matter how large the eusocial reproductive advantages are. Thus, it is also a `high risk' strategy. To see why this is the case, let us consider the most favorable scenario for eusociality: eusocial queens become immortal once the colony reaches size $m$. 

In this case, the extinction risk of a eusocial colony lies in the intermediate transition steps and reduces monotonically to zero until reaching the threshold size $m$. Note that $q = 1$ makes it fastest for a single eusocial queen to reach the eusocial threshold and thus the immortal state. There is no intermediate optimal $q$ to this end. Therefore, if the extinction probability, $p_1$, of a single queen with full eusociality ($q = 1$) is lower than that of a single solitary individual, then it is advantageous for the eusocial queen to utilize a fully eusocial strategy ($q = 1$) to minimize the risk of extinction. 

To simplify our notations, we can normalize the birth and death rates with their sum $b_i + d_i$ in the ordinary differential equations in Eq.\ref{mbp} without changing the original fixed points. Let $a_i = b_i/(b_i + d_i)$ for $1\le i \le M$. Then $a_i = a_0 = b_0/(b_0 + d_0)$ for $i < m$; $a_i = a = b/(b+d)$ for $i \ge m$. The extinction probability, $p_1$, of a single fully eusocial queen is

\begin{equation}
p_1 = \sum_{i = 1}^{m-1} (1 - a_0) a_0^{i-1}.
\end{equation}

The condition for $p_1$ to be smaller than the extinction probability of a single solitary individual $p_0 = (1-a_0)/a_0$ is

\begin{equation}
W_m(a_0) = a_0^{m-1} + a_0^{m-2} + \cdots + a_0 - 1 < 0.
\end{equation}

The above polynomial $W_m(a_0) = a_0^{m-1} + a_0^{m-2} + \cdots + a_0 - 1$ always has a unique positive root, denoted by $r_m$, in $(0,  1)$. Thus eusociality can have an advantage over solitary (in terms of lower extinction probability) if and only if the ratio of solitary reproduction $a_0 = b_0/(b_0 + d_0) < r_m$. This yields the critical solitary $R_{S}(m)$ that always ensures solitary reproduction less prone to extinction than eusocial reproduction for any given eusocial threshold $m$:

\begin{equation}
\frac{b_0}{d_0} > \frac{r_m}{1 - r_m} = R_{S}(m).
\end{equation}

For $m = 2$, $r_2 = 1$. Even so, we stress that this does not mean that eusocial reproduction is less subject to extinction than solitary reproduction for this case with $m = 2$, since this threshold $r_2$ is calculated with assuming immortal eusocial queen ($d = 0$). In fact, for eusocial queens without infinite lifespan ($d > 0$), it requires very large eusocial benefits to reduce the extinction risk than solitary reproduction does. To see why, let us recall Eq.~\ref{extp} which we derived for the simplest possible case $m = M = 2$: e.g., when the solitary reproduction $b_0/d_0 = 3$, the eusocial reproduction needs to satisfy $b/d >b_0/d_0(1 + b_0/d_0) = 12$ to guarantee a lower extinction probability than its counterpart, the solitary reproduction does. 

For $m = 3$ we find 

\[
\begin{array}{rl}
r_3  & = \frac{\sqrt{5}-1}{2}\approx 0.618,\\
 R_{S}(3) & =\frac{1 + \sqrt{5}}{2} \approx 1.618,
\end{array}
\]

We note that this critical value is the `golden ratio'. For $m = 4$,

\[
\begin{array}{rl}
r_4 & = \frac{1}{3} \left(-1-\frac{2}{\left(17+3 \sqrt{33}\right)^{1/3}}+\left(17+3 \sqrt{33}\right)^{1/3}\right) \approx 0.544,\\
R_{S}(4) & \approx 1.193.
\end{array}
\]

Because $W_{m-1}(a_0) < W_m(a_0) = (1 - a_0^m)/(1-a_0) - 2$, the unique root $r_m$ within $(0 , 1)$ for $W_m(a_0) = 0$  is monotonically decreasing with $m$ and converges to $1/2$ for large $m$ values, which implies that $R_{S}(m)\to 1$ for large $m$. As a result, as the eusocial threshold $m$ increases, it becomes more likely that the solitary reproductive ratio $b_0/d_0$ exceeds the threshold $R_{S}(m)$, making it impossible for eusociality to have a higher probability of emergence than solitary. In other words, the larger $m$, the more prone eusocial reproduction is to extinction than solitary reproduction.


\subsection{Numerical results}

\begin{figure}[htbp]
   \centering
   \includegraphics[width=\columnwidth]{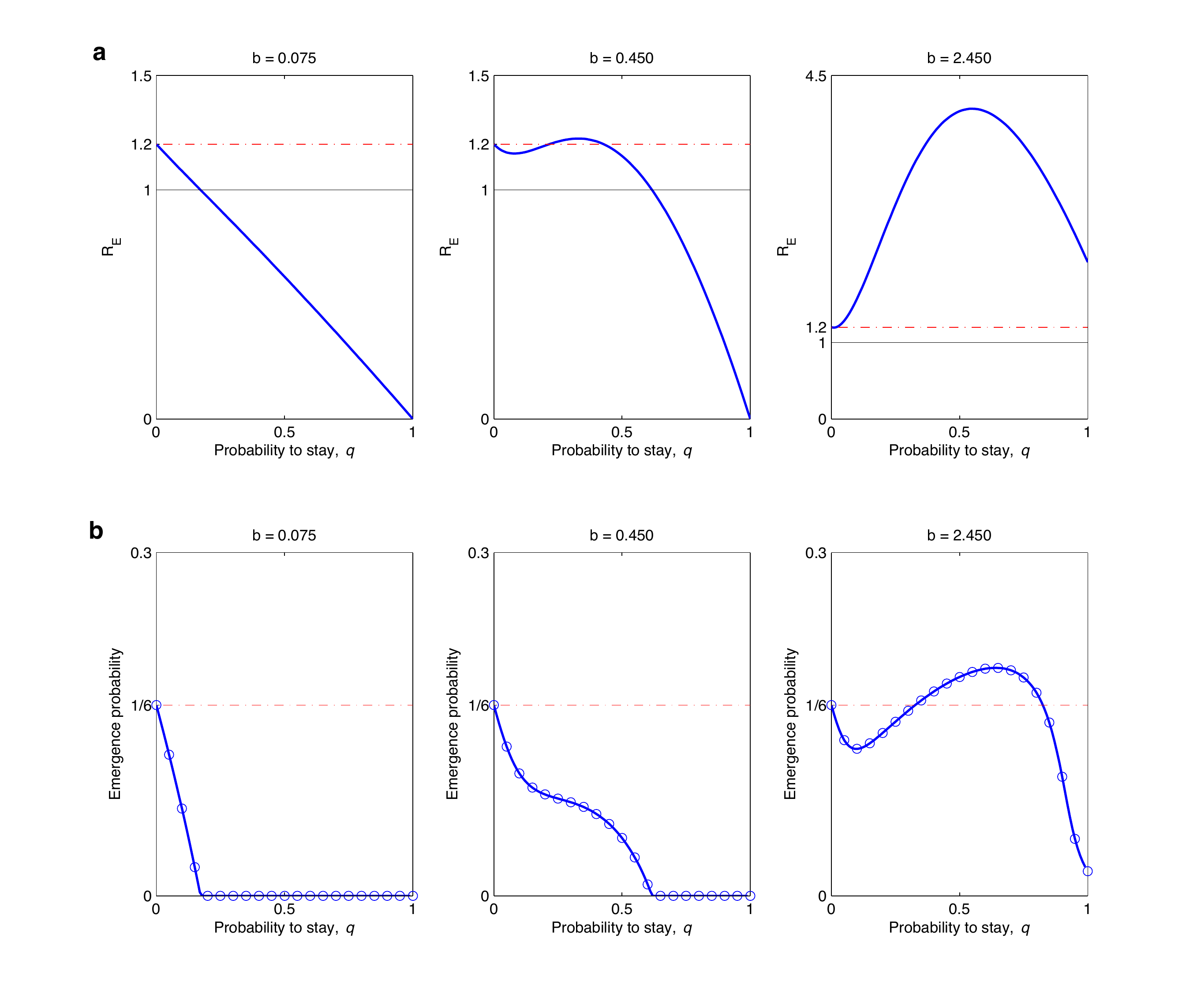} 
      \caption{Stochastic emergence of eusociality. The upper row (\textbf{a}) shows the $R_E$ of eusociality (Eq.\ref{ro}); the horizontal solid lines are the critical threshold $R_0 = 1$, while the horizontal dot dashed lines are the solitary $R_S = b_0 / d_0$. The lower row (\textbf{b}) shows the ultimate emergence probability for one single eusocial queen to establish surviving lineages; the horizontal dot dashed lines are the emergence probability of a single solitary individual, $(b_0 - d_0)/b_0$. The solid curves are the closed form evaluations of the emergence probabilities as given in Eq. (20), and the circles are results obtained by numerically solving the differential equations [Eq. (1-2)]. With increasing efficiency of eusocial reproduction ($b$), there exist an intermediate range of $q$ values such that a single eusocial queen has a higher emergence probability than a solitary individual. Relating emergence probability to $R_E$, we find that the condition for the multi-type branching process to be supercritical (\emph{i.e.}, non-zero ultimate probability for eusociality to emerge) is equivalent to requiring $R_E > 1$. Parameters: $m = 3$, $M = 100$, $b_0 = 0.12$, $d_0 = 0.1$, $d = 0.05$.
}
   \label{figexR}
\end{figure}

We present numerical results to verify our theoretical predictions above. Figure 2 shows the emergence probability and $R_E$ of eusociality as a function of the probability to stay, $q$, with increasing eusocial birth rates $b$. For a small eusocial advantage, the emergence probability decreases monotonically with $q$ until reaching certain extinction. As the eusocial advantage further increases, the emergence probability has a peak for intermediate values of $q$. Eusociality has a greater emergence probability than solitary with a sufficiently large eusocial advantage and intermediate values of $q$. As also shown in Fig. 2, eusociality will certainly go extinct if the eusocial $R_E$ falls below one. The eusocial $R_E$ being greater than the solitary $R_S$ is not sufficient to guarantee that eusociality has a higher emergence probability, although eusociality grows faster in the deterministic sense. Furthermore, the critical $b$ based on the comparison of extinction probabilities is much larger than the one obtained by comparing eusocial $R_E$ versus solitary $R_S$.

\begin{figure}[htbp]
   \centering
   \includegraphics[width=\columnwidth]{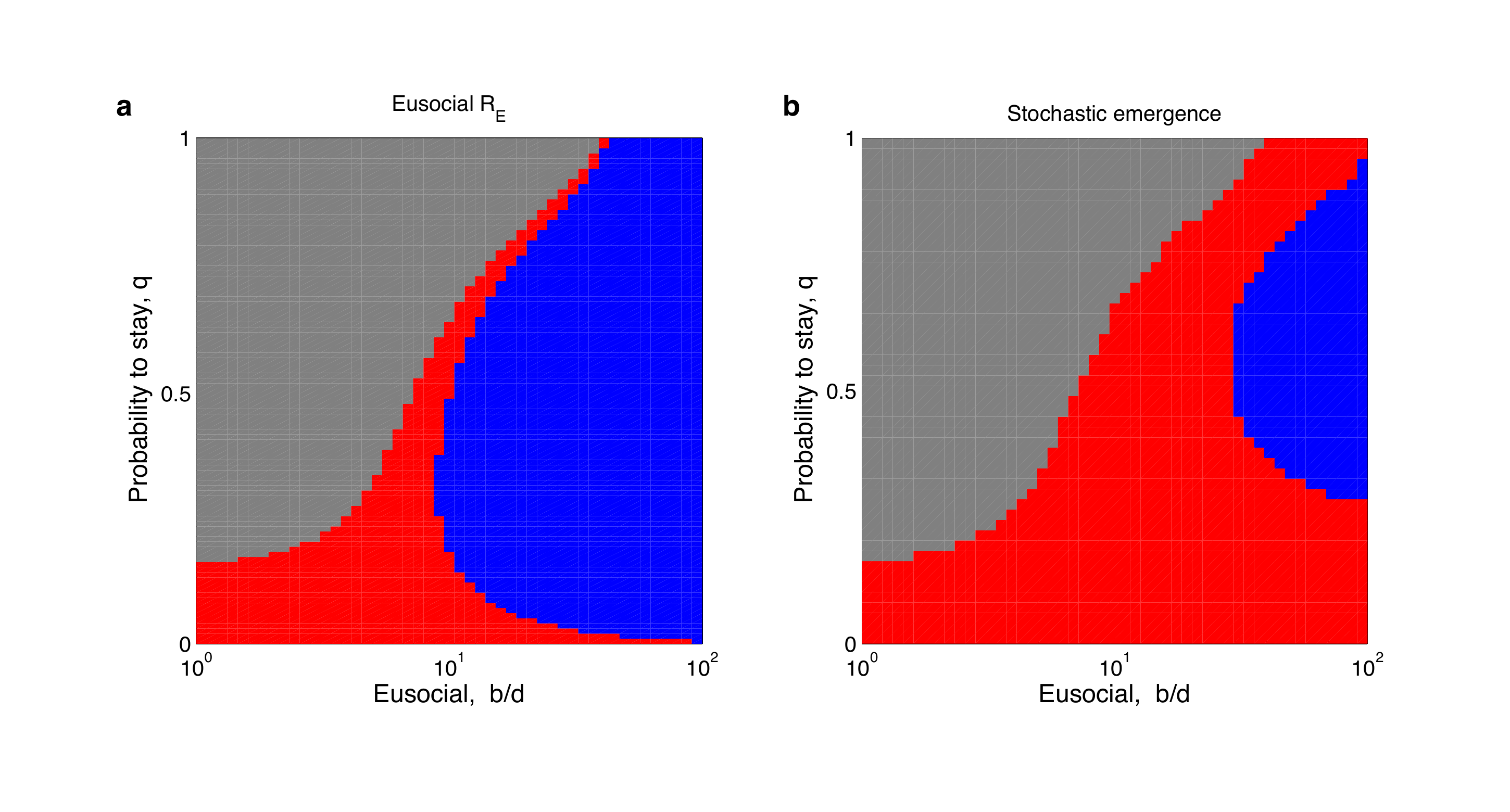} 
      \caption{Stochastic emergence and $R_0$. The plot compares the critical conditions for eusociality to have an advantage over solitary, in terms of higher $R_0$ and higher stochastic emergence respectively, across the parameter space $(b/d, q)$. In the left panel (\textbf{a}), the grey-shaded area denotes the eusocial $R_E < 1$, in the red-shaded area, eusociality has a lower $R_0$ than solitary, and in the blue-shaded area, eusociality has a higher $R_0$ than solitary. In the right panel (\textbf{b}), the grey-shaded area denotes definite extinction of eusociality, in the red-shaded area, eusociality has a lower emergence probability than solitary, and in the blue-shaded area, eusociality has a higher emergence probability than solitary. We show that the branching process is supercritical, which is equivalent to $R_E > 1$. Therefore, the grey-shaded areas in both panels are identical. The critical eusocial advantage, $b/d$, above which eusociality has a higher emergence probability than solitary, is much higher than the one above which eusocial $R_E$ is greater than solitary $R_S$. Parameters: $m = 3$, $M = 100$, $b_0 = 0.12$, $d_0 = 0.1$, $d = 0.05$.
}
   \label{figqb}
\end{figure}

Figure 3 plots the comparison of the two critical conditions for eusociality to have an advantage over solitary: (a) a higher $R_E$ across the entire parameter space $(b/d, q)$, and (b) a lower extinction probability. It is shown that the supercritical condition for a single eusocial queen to emerge with nonzero probability is exactly equivalent to requiring the eusocial $R_E >1$ as given in Eq.\ref{ro}. Note that the grey regions in the two panels of Fig. 3 are identical. The critical eusocial reproduction advantages for eusociality to emerge at a higher probability than solitary are much larger than eusocial $R_E$ simply exceeding the solitary $R_S$ (note the blue region in the right panel is much smaller, compared with the left panel in Fig. 3). 

This result can be intuitively understood as follows. Starting with a single founder queen, each surviving eusocial colony is likely to go through intermediate stepwise transitions until reaching the largest possible size $M$. Meanwhile, unlike solitary individuals, when a eusocial colony dies, it inevitably results in the simultaneous loss of all individuals belonging to that colony; much reproductive potential of these colonies with intermediate sizes is thus being wasted. As a consequence, much larger reproductive advantages are needed to offset such inefficiency of eusocial reproduction with respect to averting the risk of extinction. 

\begin{figure}[htbp]
   \centering
   \includegraphics[width=\columnwidth]{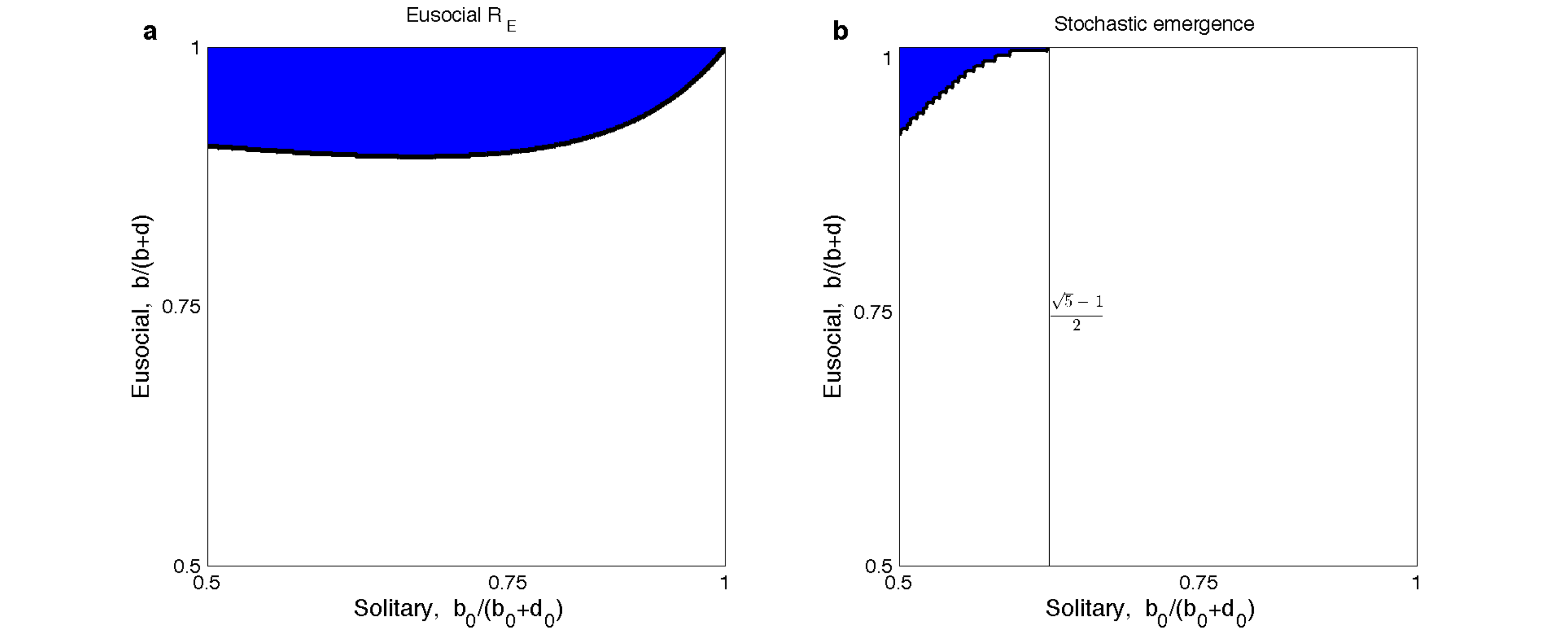} 
      \caption{Eusocial reproduction may be no better than solitary at averting the risk of extinction. The left panel (\textbf{a}) shows the region where eusocial $R_E$ is larger than solitary $R_S$. The right panel (\textbf{b}) shows the region where eusociality has a greater chance of emergence than solitary. Eusociality can have a greater probability of emergence than solitary only when the ratio of solitary reproduction $b_0/(b_0 + d_0)$ is less than a critical threshold, $r_m$. For the eusocial threshold $m = 3$, $r_m = \frac{\sqrt{5} - 1}{2}$,  $r_m$ decreases with $m$ and converges to $1/2$ for large $m$. In contrast, eusociality can always have a larger $R_0$ than solitary for any given ratio of solitary reproduction $d_0/(b_0 + d_0)$. Parameters: $m = 3$, $M = 100$.}
   \label{figse}
\end{figure}

Figure 4 depicts the parameter region of reproductive rates where eusociality has a higher $R_E$ (left panel) and has a higher emergence probability (right panel) than solitary $R_S$, respectively. The numerical result demonstrates that while it is always possible for eusociality to have a larger $R_0$ than solitary, it is not always possible for eusociality to have a higher emergence probability than solitary. Specifically, for a eusocial threshold $m = 3$ and when the ratio of solitary reproductive rates, $b_0/(b_0 + d_0) > (\sqrt{5} - 1)/2$, eusocial reproduction always imposes a higher risk of extinction to the founding queen, regardless of how advantageous eusocial reproduction may be. It is worth noting this is the ``golden ratio'' which ensures the superiority of solitary reproduction in averting the risk of extinction. For a wide range of solitary reproduction rates ($R_0 > (1 + \sqrt{5} )/2\approx  1.618$), eusociality is always more prone to extinction than solitary. Furthermore, as shown before, the critical ratio $r_m$, of solitary reproduction rates precipitately plunges with large $m$. For instance, $m = 5$, $r_m\approx  0.519$. This suggests that even if solitary reproduction is only slightly supercritical, a large eusocial threshold (\emph{e.g.}, $m\ge 5$) is the barrier to the emergence of eusociality, making eusocial reproduction much more vulnerable to the risk of extinction. These results are particularly useful in understanding the rarity of eusociality.

\subsection{Model Extension I: Facultative vs Obligate Eusociality}

In the previous section, we demonstrated that eusocial strategies have the greatest emergence probability with intermediate values of $q$, suggesting that facultatively eusocial strategies -- those where all daughters are capable of individual reproduction -- might be the most successful in the earliest stages of social evolution. These calculations were performed under the assumption that there is a uniform probability to stay, $q$, for any colony size $1\le i < M$. However, it is also plausible that $q$ may depend on the colony size $i$. For simplicity, we consider a step function for the $q$ profile: $q_i = q_0$ for $i < m$, and $q_i = q_1$ for $i\ge m$. 

\begin{figure}[htbp]
\centering
   \includegraphics[width=\columnwidth]{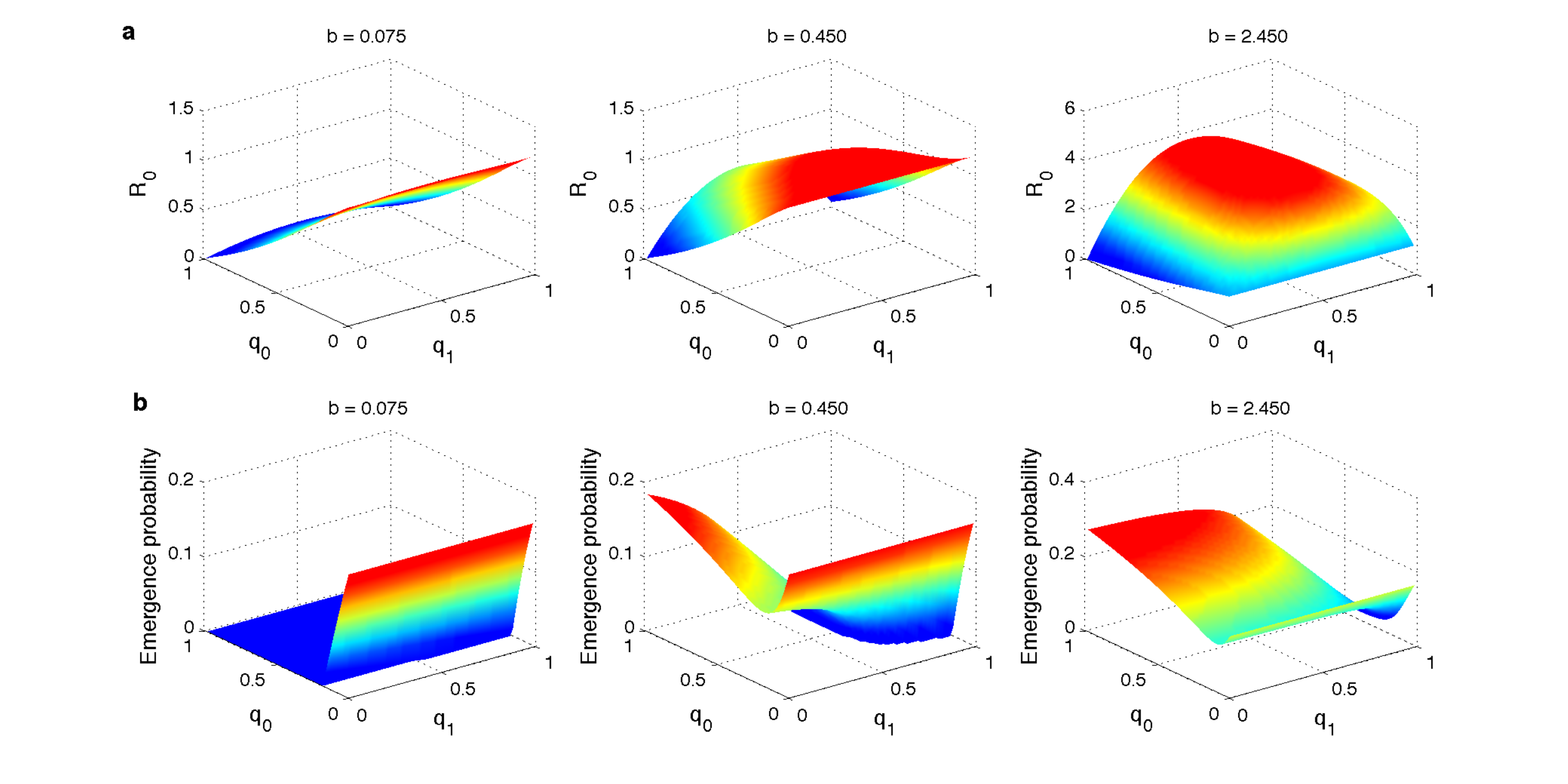}
      \caption{Optimum probability to stay. We consider a step function for the probability for offspring to stay in the nest, $q_i$, that depends on the colony size $i$:  $q_i = q_0$ for $i < m$, $q_i = q_1$ for $i \ge m$, and $q_M = 0$. The upper row (\textbf{a}) shows the eusocial basic reproductive ratio $R_E$ as a function of $(q_0, q_1)$ with increasing eusocial reproduction rate $b$. 
The lower row (\textbf{b}) shows the emergence probability as a function of $(q_0, q_1)$ with increasing eusocial reproduction rate $b$.
For small $b$ values, $q_i = 0$ maximize both $R_E$ and emergence probability. For large $b$ values, $q_0 = 1$, and $q_1 = 0$ maximize emergence probability, while $q_0 = 1$, and intermediate $q_1 \in (0,1)$ maximize $R_E$. Given large eusocial advantages, this result shows that obligate eusociality is optimal for small colony sizes; namely, it is always optimal for a queen to produce workers first followed by a burst of reproductives sent out to establish new colonies. All parameters as in Fig.~\ref{figexR}.}
   \label{qf}
  \end{figure}

We ask which $q_0$ and $q_1$ values maximize the basic reproductive ratio, $R_E$, as well as the emergence probability, $z_1$. It is straightforward to adapt the previously derived formula for $R_E$ (Eq. 16) and $z_1$ (Eq. 20) for this extended case. Figure~\ref{qf} plots the eusocial $R_E$ and emergence probability $z_1$ as a function of the parameter space $(q_0, q_1)$, with increasing eusocial reproduction rates $b$, respectively. 

For small $b$ values, solitary reproduction (\emph{e.g.}, $q_i = 0$) is most resilient against extinction and has the largest $R_0$ value. For sufficiently large $b$ values, $q_0 = 1$, and intermediate $q_1 \in (0,1)$ maximize $R_E$ while $q_0 = 1$, and $q_1 = 0$ maximize the emergence probability. Thus, our results suggest that facultative eusociality can emerge in conditions where there is a large reproductive payoff and small colony sizes, but that if a queen is capable of producing an obligate worker caste, this is the most advantageous strategy. In other words, while both facultative and obligate eusociality are capable of emerging (given the appropriate conditions), it is optimal for a queen to produce an obligate worker caste first, followed by reproductives in order to maximize the reproductive potential of the colony.

\subsection{Model Extension II: Sigmoid Functions of Eusocial Reproduction}
It is not implausible that ecological benefits begin to accrue in eusocial colonies of small size below the threshold $m$. Without loss of generality, therefore, we use the following sigmoid functions to characterize the ecological parameters governing eusocial reproduction:
\begin{eqnarray}
b_i & =  & b(i)  =  b_0 + (b - b_0)\frac{(i-1)^{\alpha_b}}{(i-1)^{\alpha_b} + (m_b - 3/2)^{\alpha_b}} ,\\
d_i & = & d(i) =  d_0 + (d - d_0)\frac{(i-1)^{\alpha_d}}{(i-1)^{\alpha_d} + (m_d - 3/2)^{\alpha_d}} ,\\
q_i & = & q(i) =  1 - \frac{(i-1)^{\alpha_q}}{(i-1)^{\alpha_q} + (m_q - 3/2)^{\alpha_q}},
\end{eqnarray}
where $\alpha_b$, $\alpha_d$, and $\alpha_q$ ($m_b$, $m_d$,  and $m_q$) are the sigmoid coefficients (thresholds) for birth rate $b_i$, death rate $d_i$, and probability to stay $q_i$, respectively. 

\begin{figure}[htbp]
   \centering
   \includegraphics[width=\columnwidth]{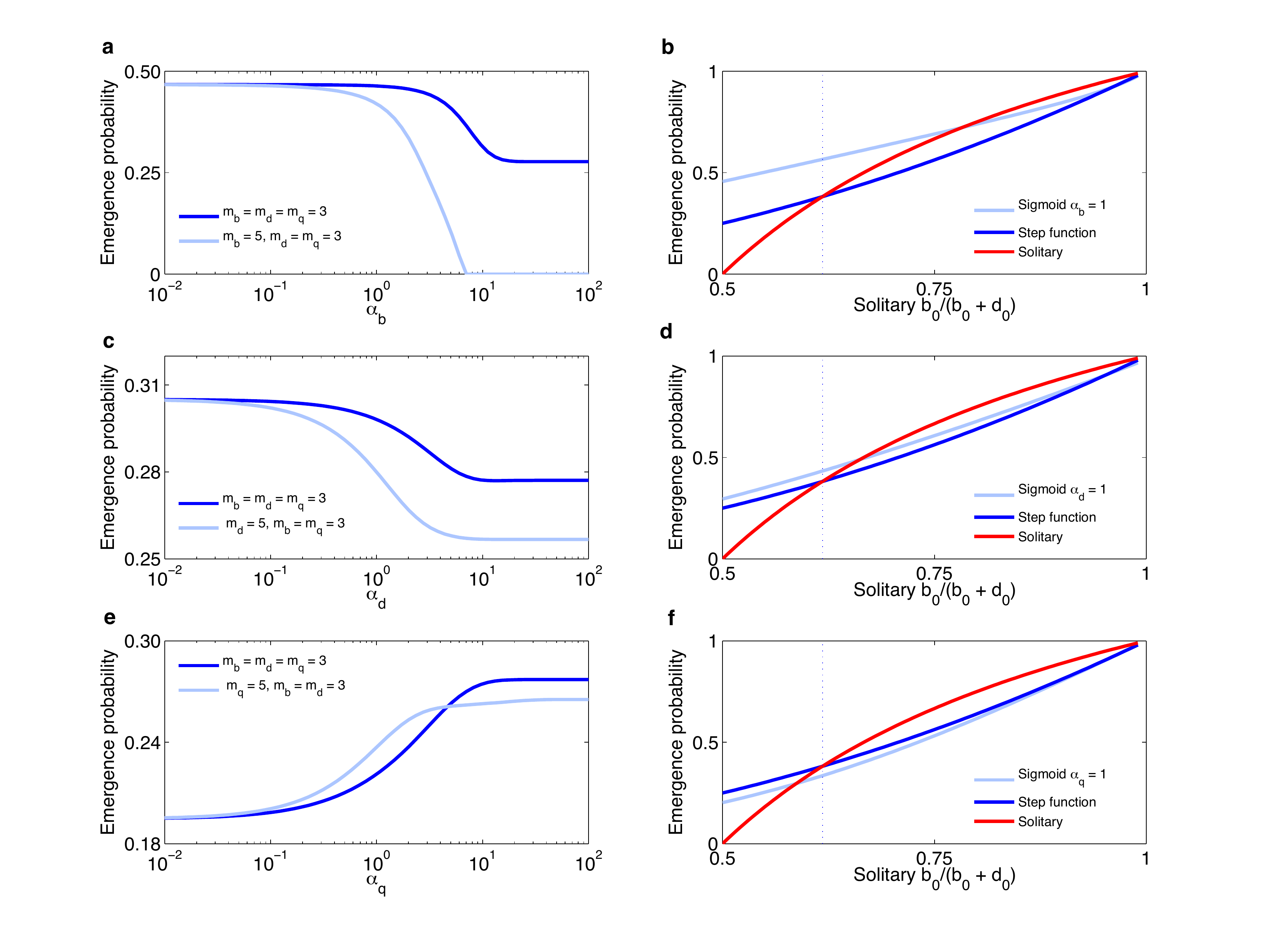} 
      \caption{Sigmoid functions. Panels \textbf{a}, \textbf{c}, and \textbf{e} show the emergence probability of a single eusocial queen as a function of the sigmoid coefficients of eusocial birth rate, $\alpha_b$, eusocial death rate, $\alpha_d$, probability to stay, $\alpha_q$, respectively. Panels \textbf{b}, \textbf{d}, and \textbf{f} plot the emergence probability of a single eusocial queen versus a solitary individual as a function of the normalized solitary reproduction rate $b_0/(b_0 + d_0)$. Vertical lines mark the golden ratio, $(\sqrt{5} - 1)/2$, as shown in Fig. \ref{figse}b. Different from step functions, assuming sigmoid functions for eusocial birth and death rates confers ecological benefits for the colony size below the threshold $m$, thus promoting the emergence of eusociality. In contrast, assuming a step  function for the probability to stay optimalizes the chance for eusociality to emerge than any sigmoid function. Under sigmoid functions, there still exist critical solitary reproduction rates such that no parameters of eusocial reproduction can lower the extinction probability ($m\ge 3$). Parameters: $M = 100$, $m_b = m_d = m_q =  m = 3$, $b = 2.45$, \textbf{a}, \textbf{c}, and \textbf{e}: $d = 0.05$, $b_0 = 0.12$, $d_0 = 0.1$, \textbf{b}, \textbf{d}, and \textbf{f}: $d = 0$, $d_0 = 0.1$.}
 \label{sigm}
\end{figure}

Note that parameters $\alpha_b$, $\alpha_d$, and $\alpha_q$ control the steepness of the sigmoid functions, and for large values, the step functions with the thresholds $m_b$, $m_d$, and $m_q$ are recovered as before. With increasing the values of $m_b$, $m_d$, and $m_q$ larger colony size is required to confer substantial eusocial benefits as well as to produce reproductives, thus making it more difficult for eusociality to emerge (Figs.~\ref{sigm}a, \ref{sigm}c, and \ref{sigm}e). Different from step functions, assuming sigmoid functions for eusocial birth and death rates confers ecological benefits for small colony size below the threshold $m$, thus promoting the emergence of eusociality (Figs.~\ref{sigm}a and \ref{sigm}c). In contrast, assuming a step function for the probability to stay optimalizes the chance for eusociality to emerge, compared to the sigmoid function (Fig.~\ref{sigm}e). Under the assumption of sigmoid functions, there still exist critical solitary reproduction rates such that no parameters of eusocial reproduction can lower the extinction probability (Figs.~\ref{sigm}b, \ref{sigm}d and \ref{sigm}f). These results confirm the robustness of our above conclusion that eusocial reproduction is in general more subject to extinction than solitary lifestyle. Moreover, our work suggests further potential empirical testing aimed at characterizing the eusocial advantages using the colony size, since to the best of our knowledge there is currently lack of accurate data to this end.

\subsection{Model Extension III: Origins and Reversals of Eusociality}

\begin{figure}[htbp]
   \centering
   \includegraphics[width=\columnwidth]{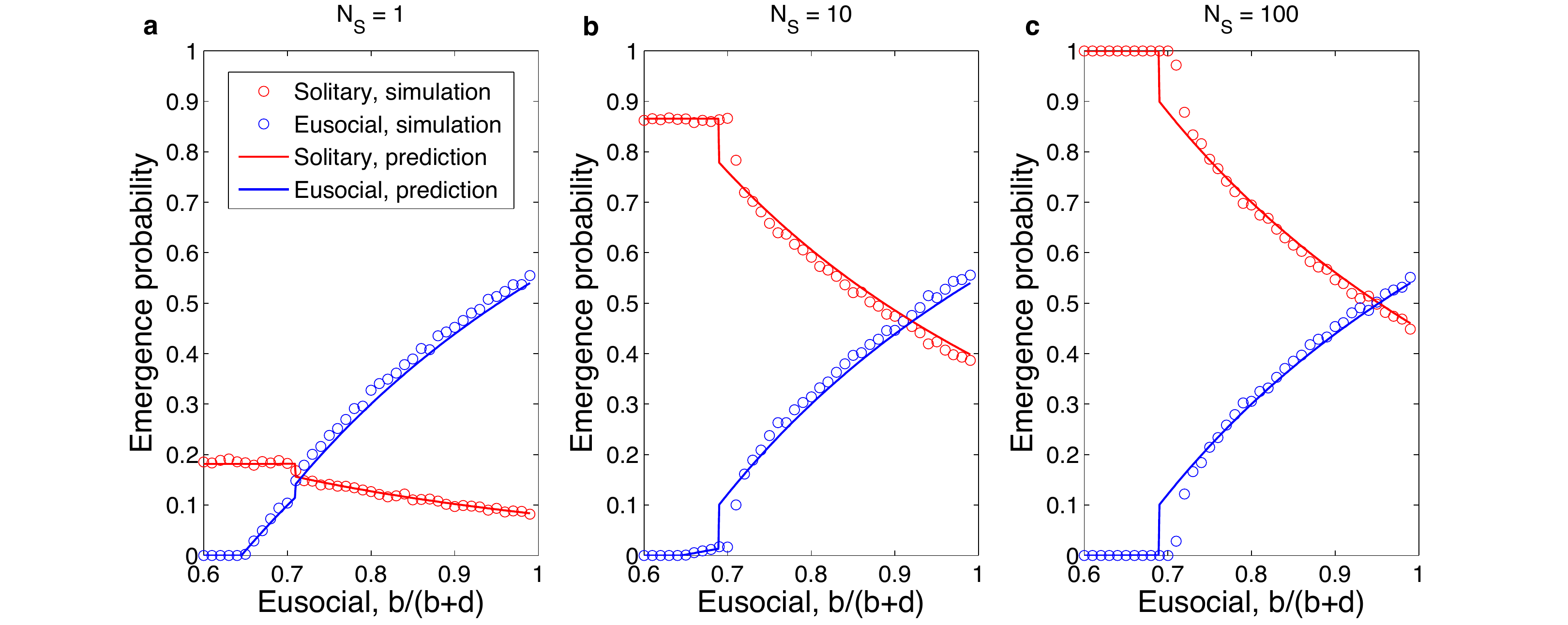} 
      \caption{Emergence of eusociality from population-size dependent branching processes. Shown are the emergence probabilities of eusociality and solitary, starting with one single eusocial individual and $N_S$ solitary individuals together ($N_S = 1, 10, 100$, respectively). The branching process is dependent on the total population size, $X$: the birth rates change as $b_i/(1 + \eta X)$ while the death rates remain constant. In this scenario, coexistence between eusocialty and solitary is impossible. There is good agreement between predictions (solid lines) and simulations (circles). Parameters: $\eta = 0.0001$, $b_0 = 0.55$, $d_0 = 0.45$, $b = 1 - d$, $q = 1$, $m = 2$, $M = 2$.
}
 \label{figdr}
\end{figure}

Let us consider the scenario where the branching process is dependent on the total population size $X$: the birth rates are rescaled by a factor $\phi =1 /(1 + \eta X)$, $b_i' = \phi b_i$, with $\eta > 0$, and the death rates remain unchanged. The parameter $\eta$ quantifies the intensity of competition and determines the upper bound of average population size. Coexistence between eusociality and solitary is impossible for $\eta > 0$. Therefore, the extinction probabilities of solitary and eusociality (denoted by $p_s$ and $p_e$) calculated before are still relevant in this scenario. In particular, we can still use the extinction probabilities calculated before ($\eta = 0$) to predict the evolutionary outcomes for small $0 < \eta \ll 1$. For $\eta = 0$, coexistence happens with probability $(1 - p_e) (1 - p_s)$, only solitary emerges while eusociality goes extinction with probability $p_e (1 - p_s)$, only eusociality emerges while solitary goes extinction with probability $p_s (1 - p_e)$, and both species go extinction with probability $p_s p_e$. Because extinction, if fated to happen, takes place in short time scales for branching processes, the co-existent state that happens with probability $(1 - p_e) (1 - p_s)$ for $\eta = 0$ will be dominated by and eventually taken over by the one having a higher $R_0$ in the case of $\eta > 0$. For solitary reproduction, $R_S = \phi b_0 /d_0$; for eusocial reproduction ($M=m=2$ in Fig. S1), $R_E = \phi^2 b b_0/ [d(\phi b_0 + d_0)]$. Substituting $\phi = d_0/b_0$ into the inequality of $R_E > 1$, we obtain the critical ratio $b/d$ above which eusociality dominates over solitary,

\begin{equation}
\frac{b}{d} > 2\frac{b_0}{d_0}.
\end{equation} 

That being said, if $b/d > 2b_0/d_0$, then the emergence probability of eusociality and solitary is $1 - p_e$ and $p_e(1 - p_s)$, otherwise being $p_s(1 - p_e)$ and $1 - p_s$, respectively. Figure~S1 shows that this approximation works well for small $\eta\ll 1$ and for different initial numbers of solitary individuals. It requires greater eusocial advantages for a single eusocial individual to take over a larger population of solitary individuals (\emph{i.e.}, larger $N_S$).

\begin{figure}[htbp]
   \centering
   \includegraphics[width=\columnwidth]{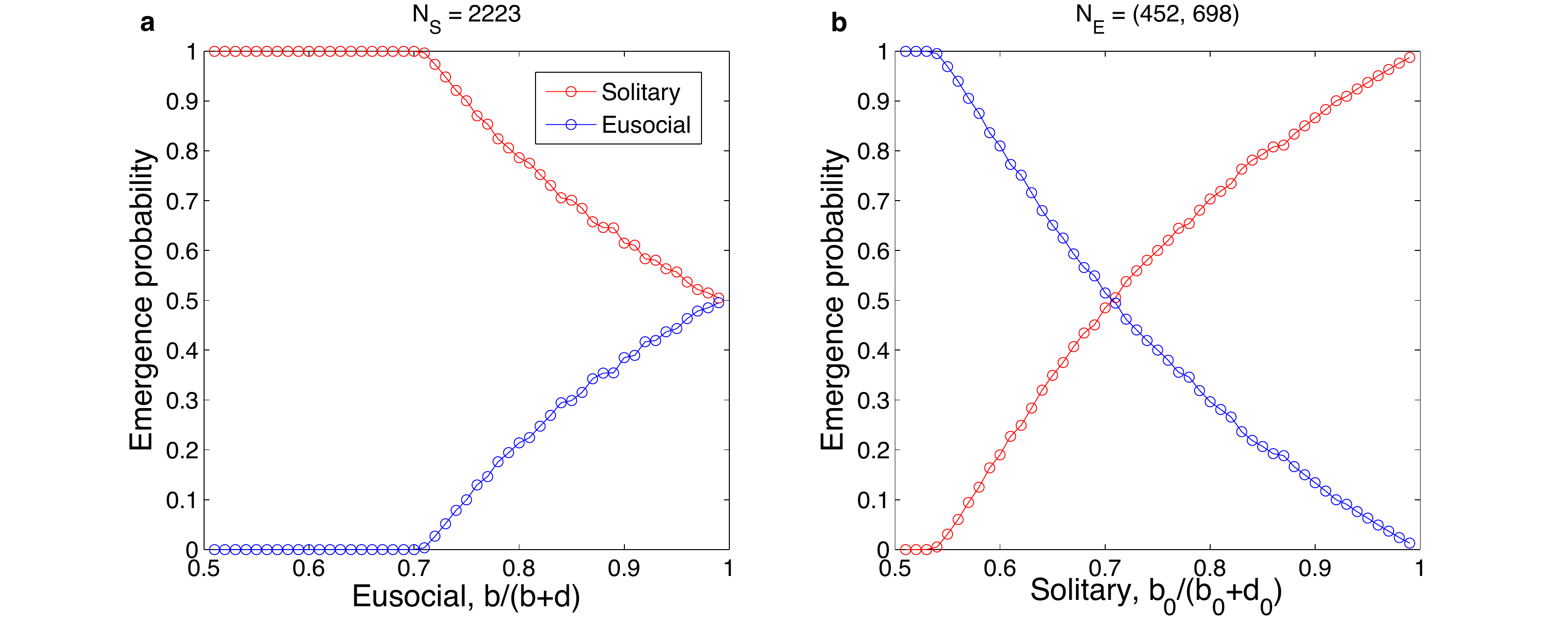} 
   \caption{Origins and reversals of eusociality. We simulate density-dependent branching processes starting with one species almost at carrying capacity and calculate how likely a single mutant can take over the entire population (\emph{i.e.}, emergence probability since coexistence is impossible). The left panel (\textbf{a}) shows the emergence probability of a single eusocial mutant in a population of $N_S = 2223$ solitary individuals. The right panel (\textbf{b}) shows the emergence probability of a single solitary mutant in a eusocial population consisting of $452$ colonies of size 1 and $698$ colonies of size 2. As a comparison baseline (neutral evolution), the probability that a single solitary individual takes over the entire homogeneous population by random drift is approximately $2\times 10^{-4}$, while that probability for a single eusocial individual is approximately $8\times 10^{-4}$. It requires very large eusocial advantages to enable successful invasion of eusociality, whereas only small solitary reproduction rates can lead to the reversal of eusociality. Parameters: $\eta = 0.0001$, $b = 1 - d$, $b_0 = 1 - d_0$, $q = 1$, $m = 2$, $M = 2$, (\textbf{a}) fixed solitary reproduction rates $b_0 = 0.55$, $d_0 = 0.45$, (\textbf{b}) fixed eusocial reproduction rates, $b = 0.7$, $d = 0.3$.
}
 \label{figinv}
\end{figure}

We further simulate the density-dependent branching processes starting with one species almost at carrying capacity and calculate how likely a single mutant can take over the entire population. The simulations show that it requires very large eusocial advantages to enable successful invasion of eusociality, whereas only small solitary reproduction rates can lead to the reversal of eusociality (online appendix, Fig. S2). These results help to explain why there are so few origins of eusociality but many subsequent losses in the social insects. 

\section{Discussion \& Conclusion}

We have calculated the extinction probabilities associated with solitary and eusocial reproductive strategies. In general, we find that solitary life history strategies are less prone to extinction than eusocial ones. Thus, solitary individuals are better at ``\emph{risk management}''. Our comparative results on the risk of extinction further support the following conclusion: the emergence of eusociality is difficult because it requires relatively large reproductive advantages over solitary strategies. Furthermore, these pronounced advantages have to arise with the production of a small number of workers. This finding is not unprecedented, and in fact, previous applications of birth-death models have suggested that the origins of eusociality might be most likely to occur in situations with strong ecological constraints favor cooperation~\citep{Aviles_EER99}.

Moreover, we derive a closed-form supercritical condition for eusociality to have a non-zero chance of emergence, if and only if the eusocial $R_E > 1$, as given in Eq.~\ref{ro}. We show that this $R_{E}$, although derived for the stochastic branching process, is exactly the same as the one obtained by the deterministic model in~\cite{Nowak_Nature10}. For eusociality to have a lower risk of stochastic extinction than solitary, much higher reproductive advantages must exist than for eusocial queens to simply have a larger $R_{E}$ than solitary females $R_{S}$ (Fig. 3). Provided sufficiently large reproductive advantages, eusocial queens can always have a larger $R_{0}$ than solitary, whereas only when the ratio of solitary reproduction is lower than a critical value, $b_{0}/(b_{0}+d_{0}) < r_{m}$, can eusociality have a lower risk of extinction than solitary (Fig. 4). In other words, eusociality is more vulnerable to the risk of extinction, even though eusocial reproduction can be very advantageous and solitary reproduction is only slightly supercritical. This is an important new insight arising from the present work that helps explains the rarity of eusociality.

In the relatively few eusocial species where the reproductive benefits have been studied, eusociality does appear to confer both higher reproductive output as well as increased offspring survivorship, even with very small colony sizes. Examples of greater reproductive payoff for social species are rare, but evidence from some socially polymorphic species suggest that social colonies do indeed have greater productivity than solitary nests~\citep{Smith_BES07} and can produce more reproductives than solitary species (2-3 reproductive females in solitary \emph{L. albipes} nests versus 12-20 reproductive females in eusocial \emph{L. albipes} nests; Cecile Plateaux-Quenu, \emph{personal communication}
~\citep{Plateaux-Quenu_ACIS93,Plateaux-Quenu _IS00}). Furthermore, nests with increased social complexity often have a higher average productivity per female than solitary nests and/or may be less likely to fail than solitary nests (reviewed in~\citep{Andersson_ARES84, Smith_BES07, Rehan _IS10, Rehan _BLS11}). 

One of the oft-cited examples of the benefits of social living is that social species are able to reduce their risk of parasitism and predation~\citep{Lin_QRB72,Evans_BS77}, which should lead to higher rates of offspring survivorship. For example, \emph{Xylocopa sulcatipes} females are capable of producing both solitary and social nests. Social nests contain a secondary female that acts as a guard at the nest entrance, and her presence significantly reduces the frequency of nest usurpation by other, conspecific females in environments with limited nest-site availability~\citep{Stark_E92}. Other studies have demonstrated similar benefits in small carpenter bees (\emph{Ceratina} spp., ~\citep{Sakagami_IS77, Rehan _IS10, Rehan _BLS11}). 

Our initial analyses were based on a model of eusocial behavior where offspring dispersal rates were not allowed to vary over the lifespan of a colony. This approximates a facultatively eusocial colony where all daughters are capable both of reproducing on their own or remaining in the nest as helpers. In this model, facultative eusociality is still capable of emerging with sufficient reproductive payoff and intermediate values of $q$. These behavioral strategies are often observed in nature. For example, some halictid bee species are capable of producing either reproductive daughters or workers in the first brood depending on their local environment~\citep{Yanega_PNAS88,Packer_BES90,Plateaux-Quenu_ACIS93,Soucy_Evo02,Cronin_IS03,Field_CB10}. Other species, such as the halictid bee, \emph{Halictus rubicundus}~\citep{Yanega_PNAS88}, and several wasp species~\citep{Haggard_CE80,Greene_JKES84,Kolmes_JNYES86}, exhibit ``graded'' resource allocation strategies where first-brood females are often a mix of workers and reproductives.

In another set of analyses, we allowed dispersal to vary throughout the lifespan of the colony. We found that eusocial advantages are maximized when queens produce workers first, followed by reproductives (Fig. 5). This is representative of an obligate eusocial colony with a reproductive strategy where daughters initially produced are required to remain in the nest as workers, and are therefore incapable of reproducing on their own. This strategy has evolved several times within the social insects in the Apidae (honey bees, stingless bees, and bumble bees), in vespid wasps (in the Vespinae), and in the ants~\citep{Michener_hup74}. Previous work has demonstrated that among species with an annual life history, this method of resource allocation is the most beneficial~\citep{ Macevicz _BES76}.  Despite these increased benefits, obligate (\emph{i.e.} advanced) eusociality is still relatively rare in comparison to facultative (\emph{i.e.} primitive) eusociality (reviewed in~\cite{Kocher_Ap14}, suggesting that facultative strategies might be a necessary step on the route to obligate eusociality. 

Clearly, many social insect species have evolved much more elaborate forms of eusocial behavior that have very high reproductive payoffs and potentially low mortality. These species have achieved this success through a variety of different mechanisms. Regardless of these mechanisms, our model demonstrates that the evolution of social behavior is difficult and high-risk, that the initial origins of eusociality are likely to occur when small colonies can produce very high reproductive payoffs. We have also demonstrated that there is likely to be a high rate of loss of eusociality within these groups. However, once eusocial life histories have taken over the population, it is possible that more complex forms of social behavior can evolve. 

In summary, this work provides key insights into some of the ecological factors that impact the origins of social behavior by demonstrating that (at least initially) eusocial strategies are high risk with high reward.

\end{document}